\documentclass{aa}
\usepackage{graphicx,pifont}
\usepackage{epsfig}
\usepackage{psfig}
\usepackage{amssymb}
\usepackage{latexsym}
\usepackage{longtable}
\usepackage{subfigure}
\usepackage{amsmath}
\usepackage{array}

\begin{document}

    \title{Asiago eclipsing binaries program. I. V432 Aur}

\author	{A.Siviero\inst{1,2}
	\and
	U.Munari\inst{2,3}
	\and
	R.Sordo\inst{2}
	\and
	S.Dallaporta\inst{4}
	\and
	P.M.Marrese\inst{1,2}
	\and
	T.Zwitter\inst{5}
	\and
	E.F.Milone\inst{6}
          }
   \offprints{U.Munari, {\tt munari@pd.astro.it}}

\institute{
	Dipartimento di Astronomia dell'Universit\`a di Padova,
	Osservatorio Astrofisico, 36012 Asiago (VI), Italy
        \and
	Osservatorio Astronomico di Padova, Sede di Asiago, 36012
	Asiago (VI), Italy
	\and
	CISAS, Centro Interdipartimentale Studi ed Attivit\`a Spaziali
	dell'Universit\`a di Padova             
	\and
	Via Filzi 9, I-38034 Cembra (TN), Italy
	\and
	University of Ljubljana, Department of Physics, Jadranska 19, 1000
	Ljubljana, Slovenia
	\and
	Physics and Astronomy Department, University of Calgary, Calgary T2N
	1N4, Canada
        }

   \date{Received date ................; accepted date ...............}

   \abstract{
The orbit and physical parameters of the previously unsolved eclipsing
binary V432~Aur, discovered by Hipparcos, have been derived with errors
better than 1\% from extensive Echelle spectroscopy and $B$, $V$ photometry.
Synthetic spectral analysis of both components has been performed, yielding
$T_{\rm eff}$ and $\log g$ in close agreement with the orbital solution, a
metallicity [Z/Z$_\odot$]=$-$0.6 and rotational synchronization for both
components. Direct comparison on the theoretical $L, T_{\rm eff}$ plane with
the Padova evolutionary tracks and isochrones for the masses of the two
components (1.22 and 1.08 M$_\odot$) provides a perfect match and a 3.75~Gyr
age. The more massive and cooler component is approaching the base of the
giant branch and displays a probable pulsation activity with an amplitude of
$\Delta V$=0.075~mag and $\Delta$rad.vel.=1.5~km~sec$^{-1}$. With a $T_{\rm
eff}$=6080~K it falls to the red of the nearby instability strip populated
by $\delta$~Sct and $\gamma$~Dor types of pulsating variables. Orbital modeling
reveals a large and bright surface spot on it. The pulsations activity and
the large spot(s) suggest the presence of macro-turbulent motions in
its atmosphere. They reflect in a line broadening that at cursory inspection
could be taken as indication of a rotation {\em faster} than
synchronization, something obviously odd for an old, expanding star.
            }

   \maketitle

\section{Introduction}

Eclipsing binaries are a recognized high performance tool to derive
fundamental stellar parameters (like masses and radii), and as such they
play a key astrophysical role over the whole HR diagram (cf. Andersen 1991).
They also provide a {\em geometrical} distance determination, and comparison
with Hipparcos parallaxes supports this statement (e.g. Semeniuk 2000). The
more accurate the data and sophisticated the analysis, the closer the match
with Hipparcos parallaxes, usually within just a few per cent. Eclipsing
binaries also offer the possibility to {\em close the loop} with, for
example, techniques of synthetic spectroscopy. The synthetic modeling of
spectra of SB2 (i.e. lines from both components are visible) 
eclipsing binaries at quadrature (for ex. via Kurucz, MARCS,
NextGen, and other families of model atmospheres and synthetic spectra)
allows to derive fundamental parameters like temperature and surface gravity
that can be directly compared with the results of the orbital modeling.

    \begin{table}
    \caption{Heliocentric radial velocities of V432~Aur. The columns give the
    spectrum number (from the Asiago Echelle log book), the orbital phase
    according to Dallaporta et al. 2002b, the heliocentric JD ($-$2452000),
    the radial velocities of the two components and the corresponding
    errors, and the $<$S/N$>$ of the spectrum averaged over the six Echelle
    orders considered in the analysis.}
    \begin{center}
    \centerline{\psfig{file=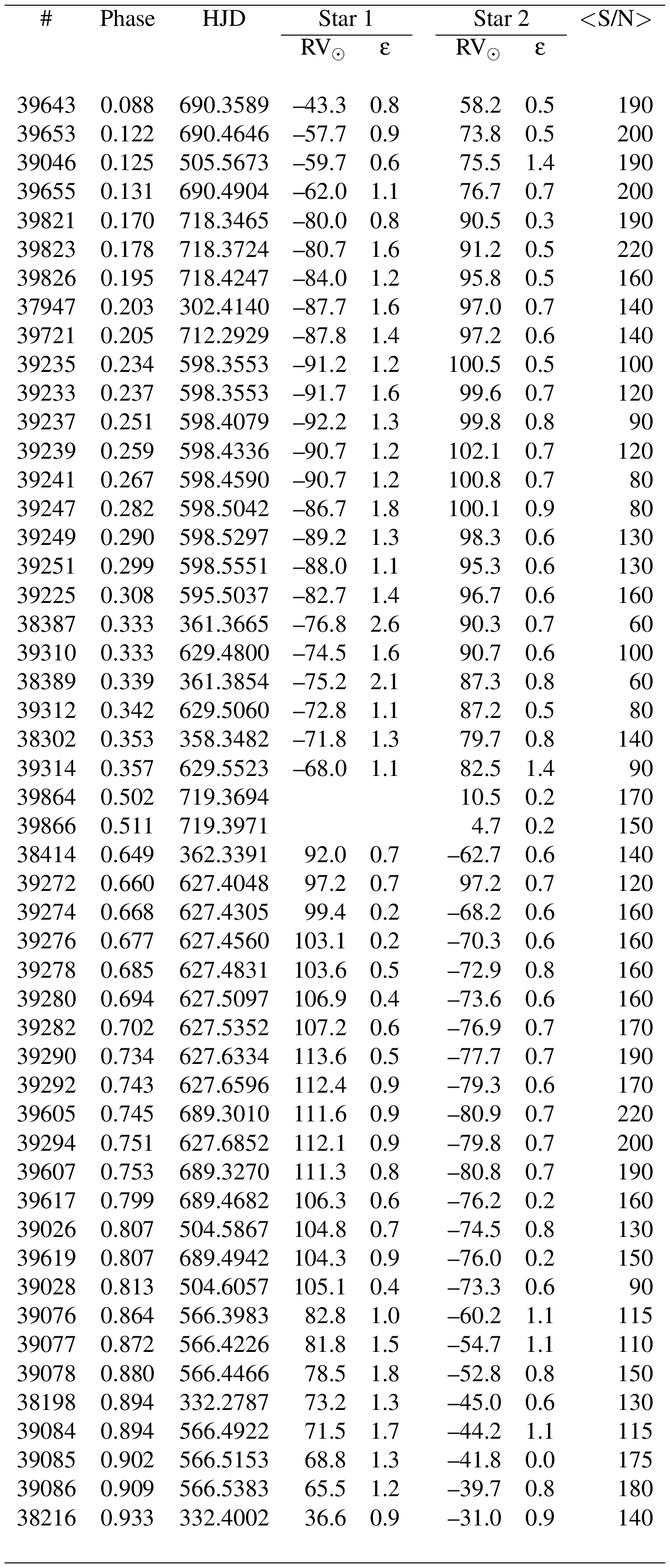,width=8.7cm}}
    \end{center}
    \end{table}

With this paper we begin a series of papers aimed to present the results
obtained on a number of SB2 eclipsing binaries selected among those missing
a solution in literature, or with improvable results. For the program stars
we have obtained accurate photoelectric or CCD photometry, generally in $B$
and $V$ bands of the Johnson system, and high S/N, high resolution
(resolving power 20000) Echelle+CCD spectroscopy over the whole
4550-9480~\AA\ range. The goal is to provide accurate orbital solutions and
stellar parameters, analyze the stellar atmospheres via synthetic spectra
techniques and derive evolutionary status and age via comparison with
evolutionary stellar models and tracks. A further goal is to provide the
proper reference against which to judge the accuracy obtainable on SB2
eclipsing binaries by the ESA's GAIA mission to be launched in 2010, a
GAIA-like orbital solution being already obtained elsewhere for the majority
of the systems considered in this series (e.g. Munari et al. 2001, Zwitter et al.
2003).

We open this series of papers with V432~Aur, a late F double lined eclipsing
binary discovered as an ``unsolved'' variable by Hipparcos. We have already
obtained accurate $B$ and $V$ photometry of V432~Aur discovering that it is a
fine eclipsing binary with a 3.1 day orbital period, and with one of the
components being an intrinsic variable (Dallaporta et al. 2002b, hereafter
Da02b). With this information derived from photometry in hand
we begun to obtain spectroscopic observations. This paper presents a
complete analysis of the combined photometric and spectroscopic data.

\section{The data}

\subsection{Photometry}

Photometric $B$ and $V$ data are those already described in detail by Da02b,
that used an Optec~SSP5 photoelectric photometer housing standard Johnson
filters. Here we augment the photometric data by an additional set obtained,
exactly in the same way, around the secondary eclipse to investigate the
long term photometric stability of the system and to check in more detail the 
reduction of intrinsic variability during the secondary eclipse.
The whole set of photometric data are available in electronic form\footnote{at {\tt
http://ulisse.pd.astro.it/Binaries/V432\_Aur/}}. In all, 1139 points in $B$
and 1574 in $V$ densely map the whole lightcurve, with a r.m.s. accuracy of
5 millimag.

    \begin{table}
    \caption{Epoch of photometric minima of V432~Aur.}
    \begin{center}
    \begin{tabular}{c c}
    \hline
    &\\
    primary & secondary\\
    &\\
    2451463.561 & 2452574.522 \\
    2451571.413 &             \\
    &\\
    \hline
    \end{tabular}
    \end{center}
    \end{table}

The photometric data reported by Da02b were reduced using $V_J$=8.15,
$B_J$=8.85 from Hipparcos Catalogue for the comparison star HD~36974. They have been now
re-reduced (and as such used in the paper and listed in the web page)
using the Tycho values $V_T$=8.259, $B_T$=9.030 transformed to corresponding
Johnson values $V_J$=8.180, $B_J$=8.871 following the Bessell (2000) transformations.

\subsection{Spectroscopy}

Spectra of V432~Aur have been secured with the Echelle+CCD spectrograph on
the 1.82 m telescope operated by Osservatorio Astronomico di Padova atop Mt.
Ekar (Asiago). A 2 arcsec slit was adopted with fixed E-W orientation,
producing a PSF with a FWHM of 1.75 pixels over the whole observing
campaign, corresponding to a resolving power close to 20000. The PSF is
measured on the night-sky emission lines. Even if uniformly illuminating the
slit, they are considered a fair approximation of the stellar illumination
mode of the slit given the typical seeing (around 2 arcsec) and the manual
guiding of the star on the slit for half an hour each exposure. Analysis of
isolated telluric absorption lines confirms the value for the resolving power.

The detector has been a UV coated Thompson CCD 1024x1024 pixel, 19 micron
square size, covering in one exposure from 4500 to 9480~\AA\ (from Echelle
order \#49 to \#24). The short wavelength limit is set by a 2~mm OG~455
long-pass filter, inserted in the optical train to cut the second order from the
cross-disperser. The wavelength range is covered without gaps between
adjacent Echelle orders up to 7300~\AA. 

    \begin{figure*}
    \centerline{\psfig{file=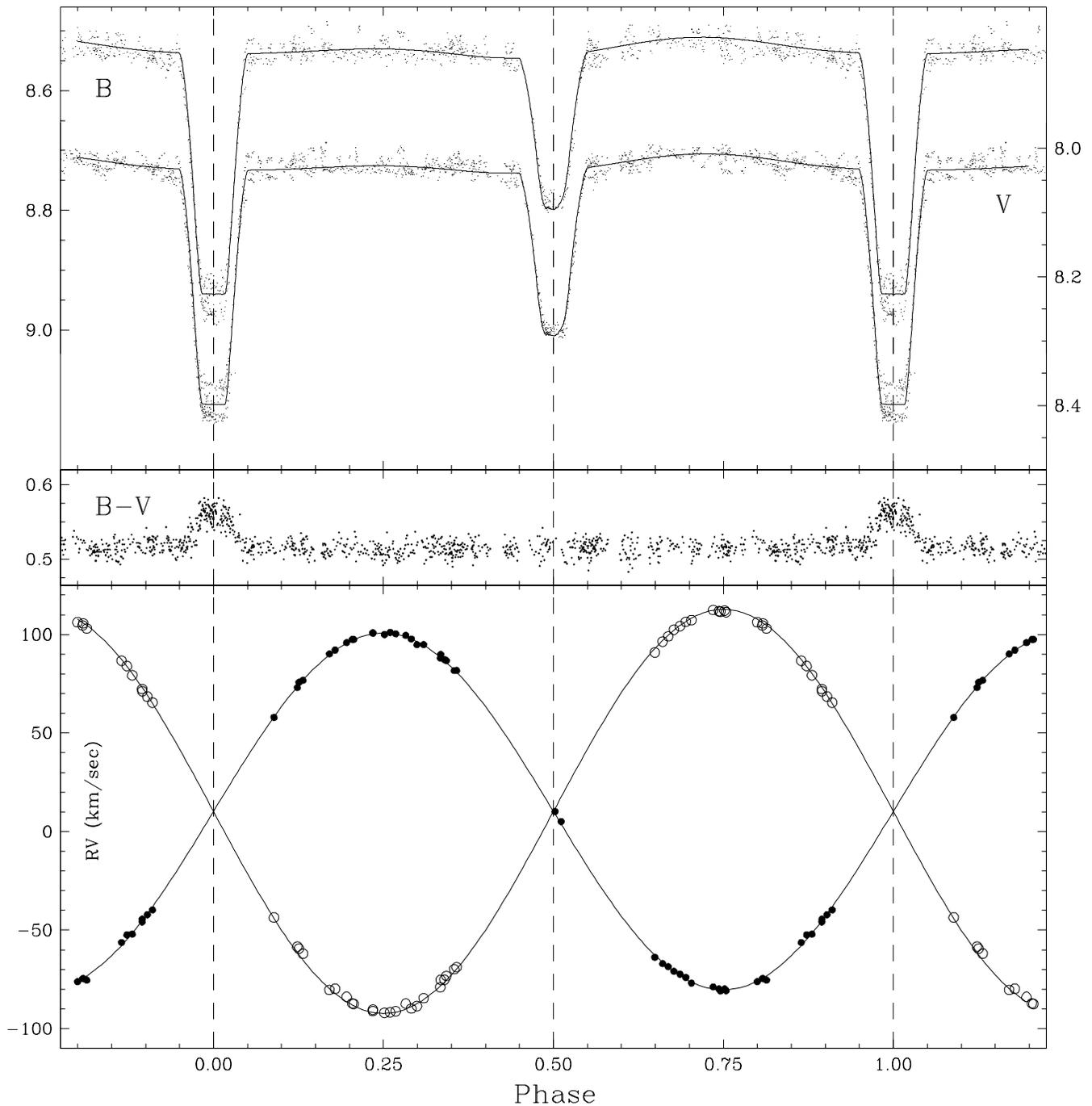,width=18.0cm}}
    \caption{The observed $V$, $B$, {\em B$-$V} and radial velocity curves of
    V432~Aur with over-plotted the orbital solution given in Table~3.}
    \end{figure*}

The exposure time has been 1800~sec for 45 spectra in Table~1 and 1200~sec
for the remaining 7, which guaranteed an excellent S/N of the recorded
spectrum while the orbital phase smearing being completely negligible
(1800~sec corresponding to 0.67\% of the orbital period).

The spectra have been extracted and calibrated in a standard fashion with
IRAF running under Linux operating system. The wavelength solution has been
derived simultaneously for all 26 recorded Echelle orders, with an average
r.m.s of 0.18~km~sec$^{-1}$. A pruned Thorium
line list has been adopted, where all lines falling into blend at our
20\,000 resolving power or approaching/reaching saturation at our 3~sec
exposure time have been deleted, leaving an average of 11 lines per order
retained in the wavelength solution.

\subsection{Radial velocities}

To derive the radial velocities, we have considered only the six Echelle
orders \#40--45 that cover the 4890--5690~\AA\ range, trimming 25\% at both
ends of each order (thus retaining the central 50\% of order). The reasons
for this are: ($a$) at these wavelengths the target spectral energy
distribution and the instrumental response reach the maximum efficiency,
($b$) the selected spectral range is packed with a great number of marked
and similar intensity FeI lines that assure easy splitting of individual
components and a great cross-correlation performance\footnote{In principle, further redder Echelle
orders could have been added in the radial velocity measurements to the aim of 
improving the accuracy of the results. We have experimented in this direction,
and the resulting gain has been marginal and unsufficient to justify the large
extra measurement effort. In fact, proceeding toward longer wavelengths the number
of stellar absorption lines per Echelle order rapidly decreases. Furthermore,
telluric absorptions (which adversely affect the accuracy of cross-correlation
measurements) begin to appear with the next redder order, number \#39, that covers
the range 5650--5840~\AA.}, ($c$) the trimmed six Echelle orders fall right at the
center of the spectrograph focal plane and CCD imaging area, where the
recorded instrumental PSF is the sharpest, ($d$) retaining only the central
50\%\ of each of the six orders avoids degradation of wavelength solution
accuracy towards the orders' ends and so maintains the accuracy of derived
radial velocities, and ($e$) due to the Echelle blaze function, the
instrument response at orders edges falls below 40\% of the peak value at
center of each order, thus producing toward order edges both a poorer S/N
and a steeper continuum (harder to normalize to unity before to run the
cross-correlation). The selected wavelength interval includes the
$\Delta\lambda$=45~\AA\ range centered at 5187~\AA\ where the highly
successful CfA Speedometers have been deriving accurate radial velocities for two
decades now, including a wealth of binaries, as described by Latham (2002, 
and references therein).

    \begin{table}
    \caption{Orbital solution for V432~Aur (over-plotted to observed data in
    Figure~1). Formal errors of the solution are given. $\dag$: adopted and fixed.
    The uncertainty in $T_2$ is estimated to be 85~K. The orbital solution fits
    only the difference ($T_1 - T_2$), derived to $\pm$8~K. The last two lines
    compare the Hipparcos trigonometric parallax (and its 1$\sigma$ error 
    interval) with the distance derived from the orbital solution.}
    \begin{center}
    \centerline{\psfig{file=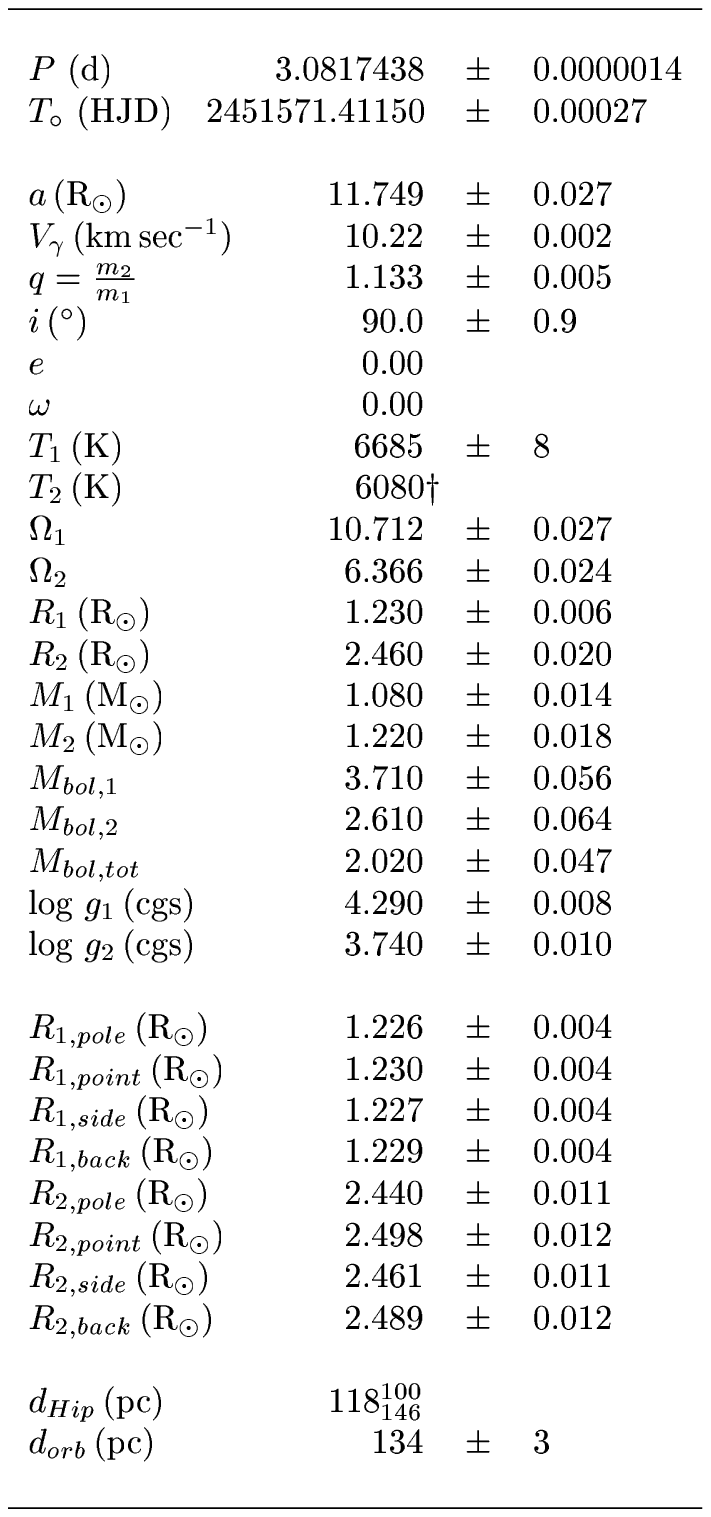,width=6.0cm}}
    \end{center}
    \end{table}

    \begin{table}
    \caption{Atmospheric parameters of V432~Aur from a $\chi^2$ fit to the
    library of synthetic Kurucz's spectra of Munari et al. (2003) computed at
    the same resolution of the Asiago Echelle spectrograph ($R$=20\,000)
    and extending over 2500$-$10500~\AA. The results from the orbital solution
    in Table~2 for $T_{\rm eff}$ and $\log g$ are given for comparison.}
    \begin{center}
    \centerline{\psfig{file=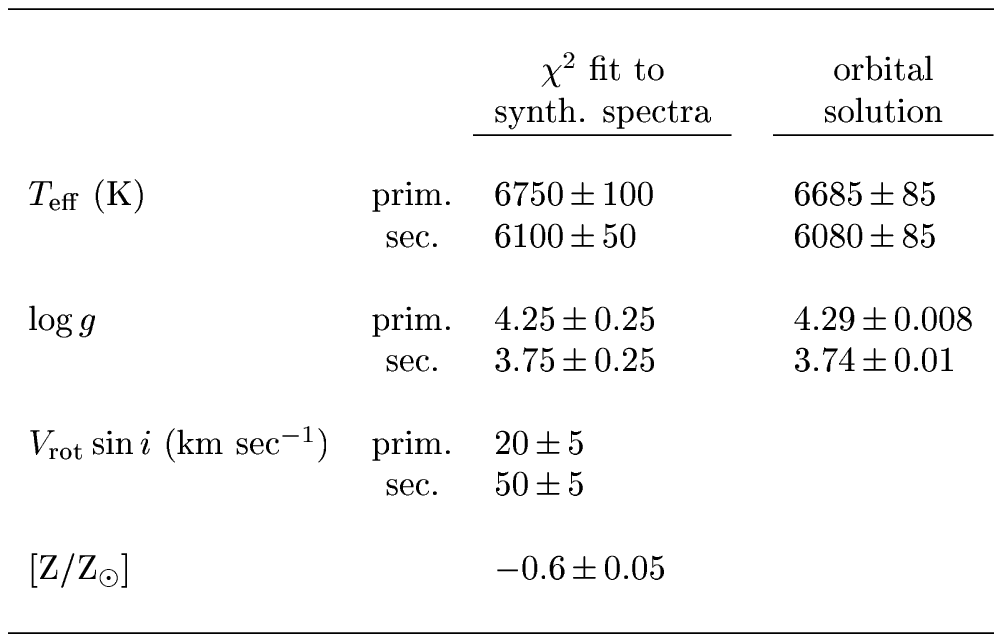,width=8.5cm}}
    \end{center}
    \end{table}

Radial velocities were measured with the two-dimensional correlation
algorithm {\tt todcor} (Zucker and Mazeh 1994). This is a multiple
correlation technique that obtains the Doppler shifts (and the intensity
ratio) of the two stellar components simultaneously. It allows for efficient
solution of even blended spectra of the two stars with an unknown intensity
ratio. The algorithm has been coded by us into a custom IRAF script which
makes use of the Fourier correlation routine {\tt xcsao} within the {\tt
rvsao} package (Kurtz and Mink 1998, Tonry and Davis 1979) and builds on a
Frotran code kindly supplied by D. Latham. The code uses template spectra of
the two stars as input. The appropriate templates have been selected among
the large synthetic spectral database computed at 20\,000 resolving power
with Kurucz's codes by Munari et al. (2003). A subsection of the synthetic
atlas, useful for F and G main sequence stars and covering the 46 Asiago
Echelle orders from 3200 to 9480~\AA\ is available electronically to
interested users\footnote{{\tt
http://ulisse.pd.astro.it/Binaries/cross\_corr/}. The grid (fits files)
covers temperatures from 7500 to 5000~K for $\log g$=4.5 (corresponding to
main sequence stars from A9 to K0), metallicities [Z/Z$_\odot$]=$-$1.0,
$-$0.5, 0.0 and rotational velocities of 0, 5, 10, 15, 20, 30, 40, 50, 75,
100~km~sec$^{-1}$. }.  The accuracy of the radial velocities from the six
selected and trimmed Echelle orders has turned out to be constant over all
six selected orders. Averaged results are summarized in Table~1.
The mean error of radial velocities is
1.05~km~sec$^{-1}$ for star~1 (the fainter of the two), and 0.66~km~sec$^{-1}$ for star~2.
The selected templates have been $T_{\rm eff}$=6500~K,
$\log g$=4.5, [Z/Z$_\odot$]=$-$0.5 and $V_{\rm rot}$=20~km~sec$^{-1}$ for star~1,
and $T_{\rm eff}$=6250~K,
$\log g$=4.0, [Z/Z$_\odot$]=$-$0.5 and $V_{\rm rot}$=40~km~sec$^{-1}$ for star~2.

    \begin{figure*}
    \centerline{\psfig{file=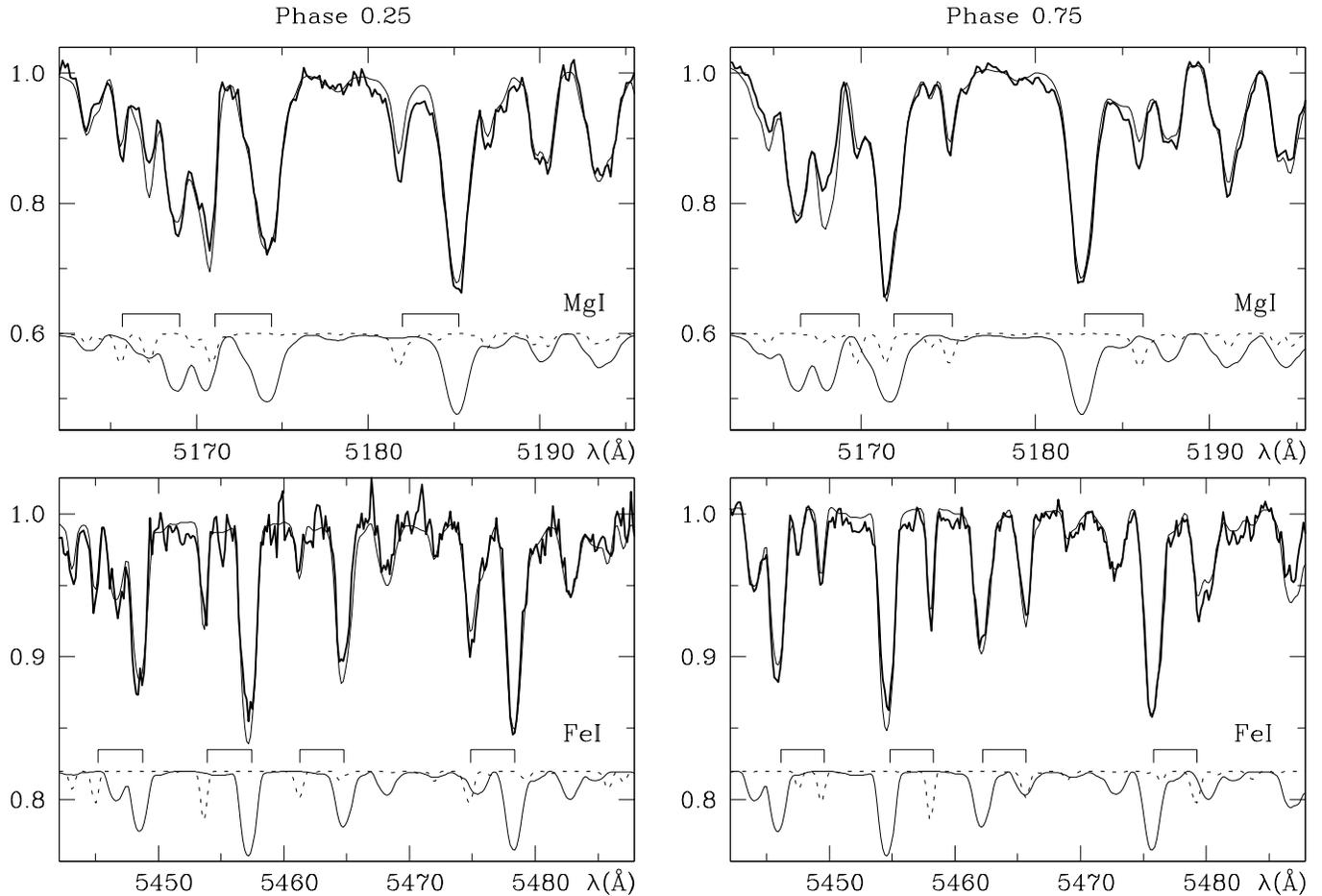,width=18.0cm}}
    \caption{Comparison between observed (thick line) and synthetic (thin line)
    V432~Aur spectra over two sample wavelength regions dominated by MgI lines
    (top) and FeI lines (bottom). Spectra at orbital phases 0.25 and 0.75 (\#
    39239 and 39294 in Table~1, respectively) show full split between the two
    components and their position interchange allow a careful check of the fit 
    accuracy. In each panel the lower curves represent (not to scale with the
    main spectra but in correct proportion between them) the contribution of
    each component of the binary to the formation of the observed spectrum at
    the given phase. The markers connect the shifted wavelengths of the same
    sample MgI (top panels) and FeI (bottom panels) lines in the spectra of the
    two components of the binary.}
    \end{figure*}

\subsection{Zeroing radial velocities}

The Cassegrain-fed Asiago Echelle spectrograph is well suited for
measurement of accurate radial velocities, with a stable and modest flexure
pattern modeled by Munari and Lattanzi (1991), and null {\em spectrograph
velocity} (in the sense introduced by Petrie 1963, i.e. the systematic
difference between the zeros of the stellar and thorium lamp wavelength
scales as due to different slit illumination and optical path).  On all
spectra of Table~1 we have checked and confirmed the null spectrograph
velocity using night-sky emission lines (selected from the list by
Osterbrock et al. 2000). The radial velocities cluster around
0.0~km~sec$^{-1}$ with a typical 0.4~km~sec$^{-1}$ dispersion.

    \begin{figure*}
    \centerline{\psfig{file=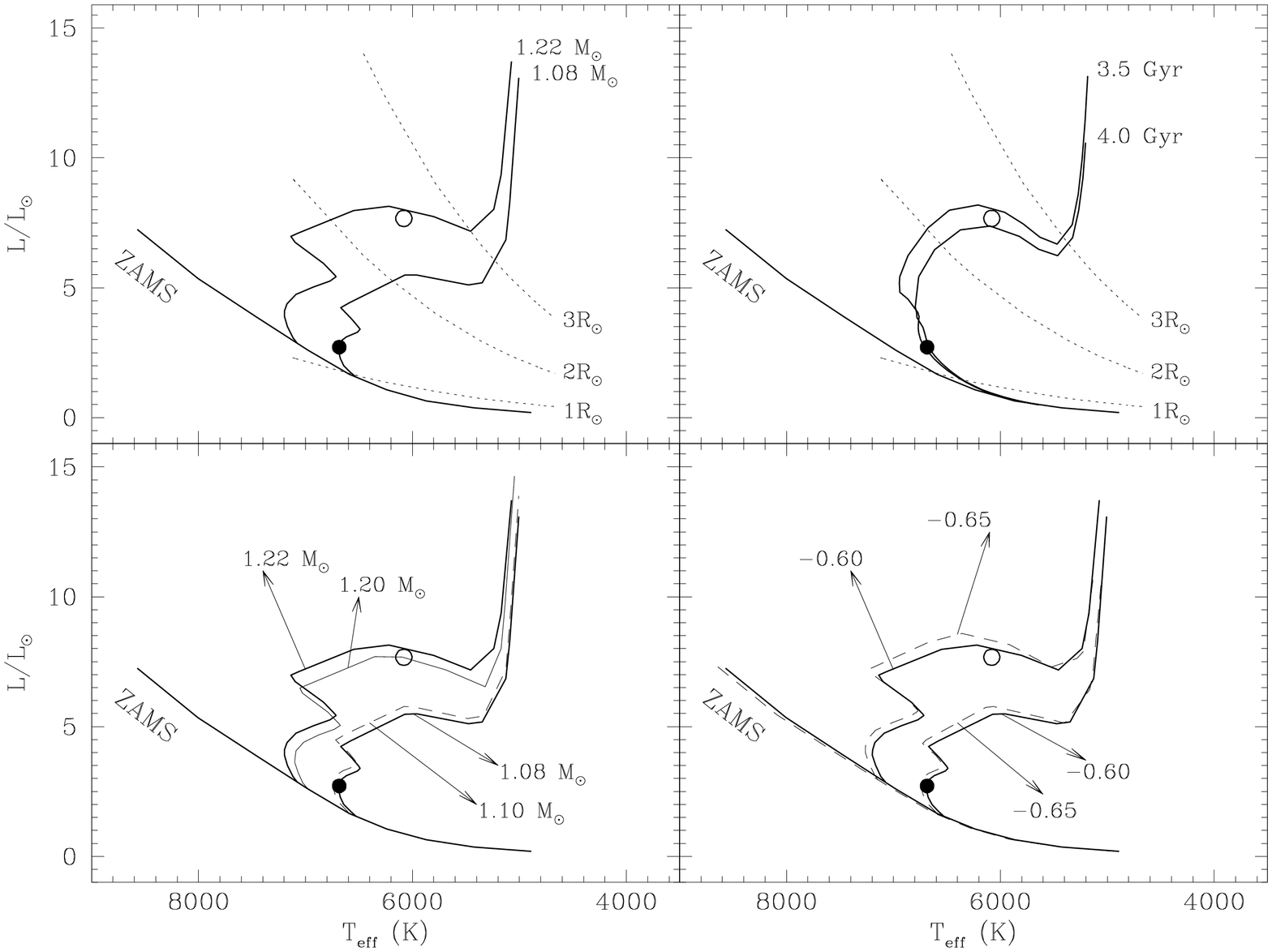,width=13.0cm,angle=270}}
    \caption{Comparison between temperature and luminosity as derived from the
    orbital solution (cf. Table~3) and those of theoretical stellar models for
    the masses of the two components of V432~Aur (1.22 and 1.08 M$_\odot$) at
    the [Z/Z$_\odot$]=$-$0.6 metallicity derived by fitting with synthetic
    spectra. The plotted theoretical Padova evolutionary tracks and isochrones
    (Fagotto et al. 1994, Bertelli et al. 1994, and follow-ups with the latest
    up-dates available via the web interface to the Padova theoretical group at
    the address {\tt http://pleiadi.pd.astro.it/}) have been derived from
    interpolation of adjacent points in the computed grid (1.00$-$1.10 and
    1.20$-$1.30 M$_\odot$, and $-$0.37 and $-$0.67 for [Z/Z$_\odot$]). The
    agreement is excellent in both position on the tracks (upper left panel) and
    on the isochrones (upper right panel), the latter supporting a 3.75 Gyr age.
    The bottom panels show how the theoretical loci move for minimal changes
    in mass and metallicity, indicating the high sensitivity of the method.}
    \end{figure*}

The spectrograph flexures can affect the results because of the observing
mode, and must be corrected for. Each 1800~sec spectrum on the star was
followed - after the 3 min required to read the science exposure and to
prepare for the thorium 3~sec exposure - by the comparison spectrum with the
telescope still tracking the star. In such a way, a shift of $\sim$18~min in
hour angle is introduced between the telescope position at mid-exposure on
the star and on the thorium lamp. To account for the flexure pattern and
possible off-center guiding of the stellar photo-center on the slit, the rich
telluric line spectrum on the reddest orders of the recorded spectra has been
cross-correlated against a template telluric spectrum, following a technique
pioneered by Griffin and Griffin (1973). The template telluric spectrum has
been obtained with exactly the same instrumental set-up on a fast rotating
early type star observed close to zenith (where the flexure pattern flattens
toward zero) with an extremely high S/N, observed so to uniformly illuminate
the spectrograph slit. Measuring the flexure pattern on Echelle orders
different from those where the stellar radial velocities are derived is
perfectly suited because the flexure pattern and the off-center guiding are
rigid shifts over the spectrograph focal plane, amounting to the same linear
dimensions over the whole CCD.

The resulting combined correction for flexures and off-center guiding has
typically been less than 1~km~sec$^{-1}$, and it is already incorporated in
the values reported in Table~1. The limited amplitude of these perturbing
effects is due to ({\em i}\,) observations obtained as close as possible to
meridian transit, ({\em ii}\,) a typical seeing matching the 2 arcsec slit
width (in the FWHM sense), and ({\em iii}\,) long enough exposures to
average out momentarily off-center guiding.

    \begin{figure*}
    \centerline{\psfig{file=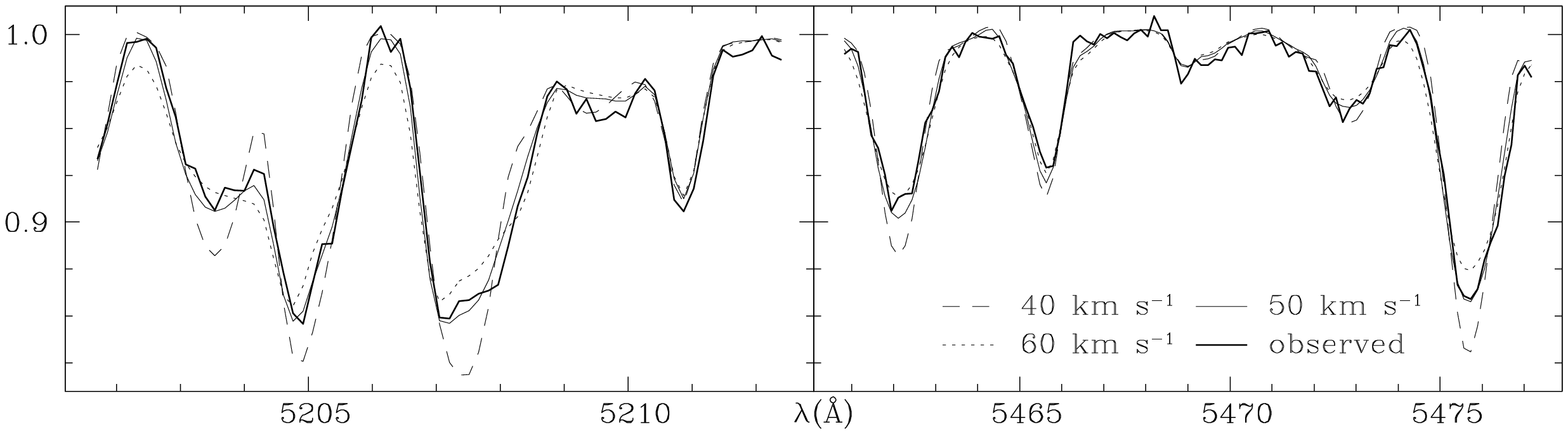,angle=270,width=18cm}}
    \caption{Rotational velocity of the secondary star in V432~Aur. The thick
     line is a short wavelength range from spectrum \#39294 at phase 0.75 (cf.
     Table~1). The thin lines, similarly to Figure~2, are the added synthetic 
     spectra (with the proper intensity ratio and velocity shift) of
     the secondary and primary component. Parameters of the synthetic spectra
     are the same as in Table~3, except that three different rotational velocities
     are explored for the secondary: 40, 50 and 60~km~sec$^{-1}$. It is evident
     how 40~km~sec$^{-1}$ (the co-rotation velocity) provides an under-estimate
     of the actual rotational velocity.}
    \end{figure*}

\section{Orbital solution}

The orbital modeling of V432~Aur has been obtained with version
WD98K93d\footnote{Latest code development are available via the web page
http://www.boulder.swri.edu/$\sim$terrell/talks/aavso2001/code/ ebdoc98.pdf}
of the Wilson-Devinney code (Wilson and Devinney 1971) as modified by Milone
et al. (1992) to include Kurucz's model atmospheres to aproximate the
surface fluxes of the two stars. The computations have been run within {\tt
Mode-2} program option. Input parameters for the first iterations (period
and epoch, semi-major axis, mass ratio, temperature of the two components,
zero eccentricity, Kopal's surface potential $\Omega_1$ and $\Omega_2$,
bolometric albedos) have been computed by hand from radial and light curves
and from spectral types of both components. They turned out to be already
quite accurate because the final achieved solution converged toward pretty
similar values. The reddening has been taken to be null because the observed
colors well match the expectation from synthetic spectral modeling (sect.4
below) and no interstellar NaI line is visible in the spectra (to a
sensitivity limit of 0.03~\AA\ equivalent to $E_{B-V} \leq $0.01 according
to Munari and Zwitter 1997 calibration).

Limb darkening coefficients have been taken from Van Hamme (1993)
interpolated for the metallicity, temperature and gravity appropriate for
the components of V432~Aur. The square root law for limb darkening has been
adopted as usual for radiative atmosphere, which is custom to assume for stars
as the components of V432~Aur. During final iterations the coefficients for the
secondary star have been slightly adjusted to achieve the closest possible
match with the light curve during primary minimum ($x_{bolo}$=0.151,
$y_{bolo}$=0.551, $x_{B}$=0.375, $y_{B}$=0.498, $x_{V}$=0.151,
$y_{V}$=0.650).

We checked for possible multiple reflection effects but, as expected given
the wide separation of the binary components, they were found negligible and
thus dropped from further modeling. Therefore only the inverse square law
illumination was considered, and the bolometric albedos were set to
1.0 and tests on the lightcurve confirmed the choice.

Finally, a gravity brightening exponent $\beta$=1.0 has been adopted
consistently with the radiative nature of the atmospheres in V432~Aur, and
co-rotation of both components has been taken in agreement with evidences
from the atmospheric spectral analysis and stability of spot longitude as
discussed below.

The resulting orbital modeling and corresponding physical quantities are
summarized in Table~3, and the model radial velocity and light curves are
over-plotted to observed values in Figure~1. In the latter no Rossiter effect
is introduced given the absence of observed radial velocities at the key
short phases around primary and secondary eclipses that could be used to
check it.

\subsection{A surface bright spot}

An evident O'Connell effect (O'Connell 1951, Davidge and Milone 1984) is
present in the V432~Aur light curve centered close to 0.75 orbital phase.
Noting ($a$) the reduction in the CaII far-red triplet line depth around
phase 0.75 in the spectra of the secondary (these are lines which usually
display emission cores in stars with active surfaces; Ragaini et al. 2003)
and ($b$) the orbital modeling indicating at phase 0.75 a model brightness
lower than observed by a few hundreds of a magnitude, we introduced a bright
spot on the surface of the secondary that brought model and observed light
curves into agreement (the model lightcurve in Figure~1 includes the effect
of the spot as well as the data in Table~3).

The spot that we found bringing model and observed light curves into
agreement is placed on the equator at 310$^\circ$ longitude on the surface
of the secondary (0$^\circ$ toward the primary, counted in the orbital
motion direction, as the convention for WD98K93 code). The spot is taken
circular, with an angular extent of 12$^\circ$ as seen from the center of
secondary star, with a 20\% increase in $T_{\rm eff}$ with respect to the
temperature of surrounding unperturbed stellar surface (7300~K). However,
different combinations of temperature, angular extent and latitude still
provide acceptable fits to the observed light curve. Our analysis is not
aiming to precise values for the spot parameters, but just to bring
attention to the fact that spot(s) are present on the surface of the
secondary star of V432~Aur.

The spot has been present during the entire period of our observations,
because the lightcurve around 0.75 orbital phase as displayed in Figure~1
comes from observations secured at various epochs, and all individual
branches display the same O'Connell effect. In particular, the longitude of
the spot appears to have remained constant during the whole observing
period, indicating both stellar co-rotation and absence of spot migration.

It could be argued that the observed O'Connell effect is instead due to a
{\em dark} spot on the other side that reduces around orbital phase 0.25 an
ellipticity effect of the secondary. We are inclined to discard this
possibility because a dark spot should not produce emission cores in the
diagnostic CaII far-red triplet lines and because the orbital modeling above
described would hardly converge in such a scenario.

    \begin{figure}[!t]
    \centerline{\psfig{file=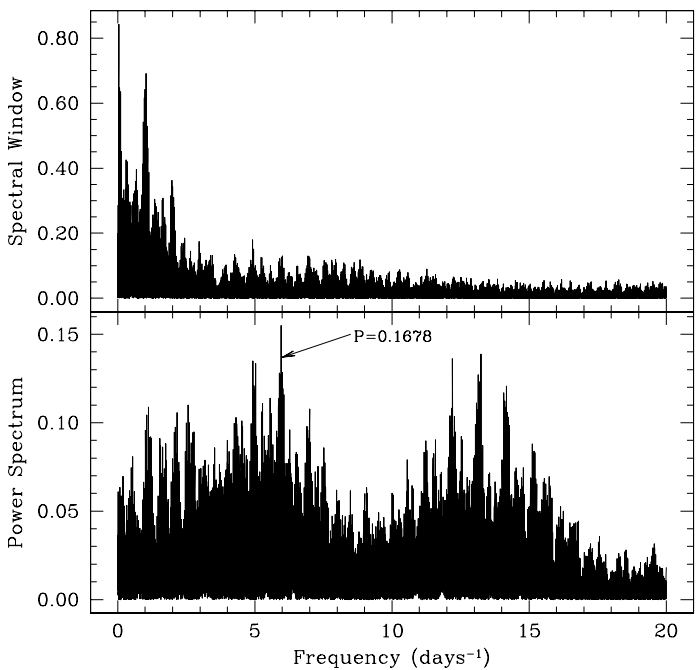,height=8.8cm}}
    \caption{Period search, for the intrinsic variable component (star~2 in 
    Table~1), on the residuals of the radial velocities from the orbital solution
    (Figure~1 and Table~2). The period ($P$=0.1678 days) corresponding to 
    the strongest peak is arrowed.}
    \end{figure}
    \begin{figure}[!t]
    \centerline{\psfig{file=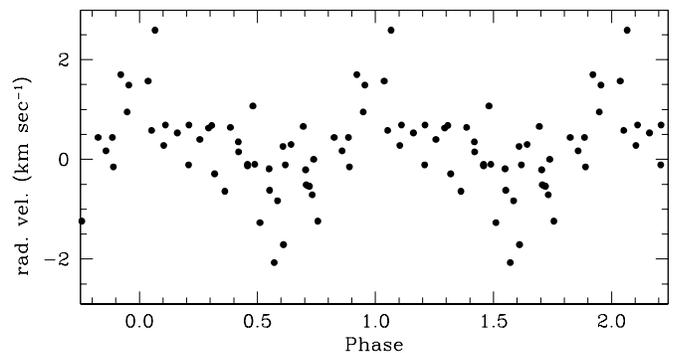,width=8.8cm}}
    \caption{Plot according to a period of $P$=0.1678 days of the residuals of 
    the radial velocities from the orbital solution for the intrinsic variable 
    component of V432~Aur (star~2 in Table~1).}
    \end{figure}

\section{Analysis of stellar atmospheres}

The spectrum of V432 Aur at primary minimum is due only to the secondary
star, and subtracting it from the spectrum at secondary eclipse would
provide the spectrum of the primary component. Unfortunately, during the
scheduled spectroscopic observing runs at the telescope, V432~Aur never
passed through the primary eclipse, and therefore it has not been possible
to obtain the isolated spectra of the individual components.

To the aim of performing an atmospheric analysis of the two components of
V432~Aur we have therefore focused on the spectra of the highest S/N around
phases 0.25 and 0.75 (cf. Table~2), where the velocity separation of the
components is maximum and the same lines from the two components are 
unbleded. Working simultaneously at phase 0.25 and 0.75 allows also 
to get rid of line superposition and strongly reinforces the robustness
of the overall fit.

To derive the basic stellar parameters ($T_{\rm eff}$, $\log g$,
[Z/Z$_\odot$] and $V_{\rm rot}$) we have performed a $\chi^2$ best match
analysis of the high S/N observed spectra around phases 0.25 and 0.75
(spectrum \#~39233, 39239, 39605, 39294 and 39607 in Table~1) against the
extensive grid of synthetic Kurucz's spectra computed by Munari et al.
(2003) for the same resolution of the Asiago Echelle spectrograph
($R$=20\,000) over the 2\,500$-$10\,500~\AA\ range (thus fully covering the
wavelength range recorded for V432~Aur). The Munari et al. (2003) synthetic
atlas covers the range 3\,500$\leq T_{\rm eff} \leq$47\,500, 0.0$\leq \log g
\leq$5.0, +0.5$\leq$ [Z/Z$_\odot$] $\leq -2.5$ (with solar relative
abundances for metals) and 0$\leq V_{\rm rot}\leq$500~km~sec$^{-1}$. The
grid steps around the values appropriate to V432~Aur are 250~K in
temperature, 0.5~dex in gravity, 0.5~dex in metallicity and 10~km~sec$^{-1}$
in rotation.

To proceed independently from the orbital solution, and thus to use the
atmospheric analysis to validate it, we had to first define the region where
the true minimum of the $\chi^2$ distribution has to be looked for, to avoid
being trapped into unphysical local minima. To accomplish this, we have
isolated from the spectrum secured at secondary eclipse (\#39684 in Table~1)
the wavelength range 8480--8740~\AA\ (dominated by diagnostic CaII triplet
and Paschen series lines). We have classified it against the spectral
atlases of Munari and Tomasella (1999) and Marrese et al. (2003) obtained
over the same wavelength range with the same instrumental set-up for the
Asiago Echelle spectrograph. We also checked the consistency of the
classification against the synthetic atlas of Munari and Castelli (2000)
that covers this wavelength region at the same resolving power of our
observations. Having determined in $\sim$6500~K the combined temperature of
the two stars, we have moved to the analytical $\chi^2$ test by fixing to
5500--7500 the temperature boundary and to $\leq$100~km~sec$^{-1}$ the
rotational velocities and imposing no restriction on gravity and
metallicity. The analysis has been limited to the same six Echelle orders
adopted for the derivation of radial velocities (cf. sect. 2.3), ready to
include additional orders if not satisfactorily converging on the first six.
This has not been necessary given the close similarity of the results among
the 6 selected orders.

The results of the atmospheric analysis are given in Table~4, where a
comparison is provided with the parameters common to the orbital solution.
Given the absolute independence of the two methods (carried out
independently by two distinct sub-groups of the authors of this paper), the
remarkable correspondence mutually reinforce the confidence in the orbital
solution and atmospheric analysis. Two small and diagnostic samples of the
4860--5690~\AA\ range investigated via $\chi^2$ test are shown in Figure~2
to show the match between observed and synthetic spectra. Inspection of the
match over the whole range does not support chemical anomalies, and
therefore the metals in V432~Aur appear to follow the relative Sun
proportions. If chemical anomalies should have emerged, it would have been required
to run a devoted atmospheric analysis (instead of the $\chi^2$ comparison to
a pre-computed grid).

    \begin{figure}
    \centerline{\psfig{file=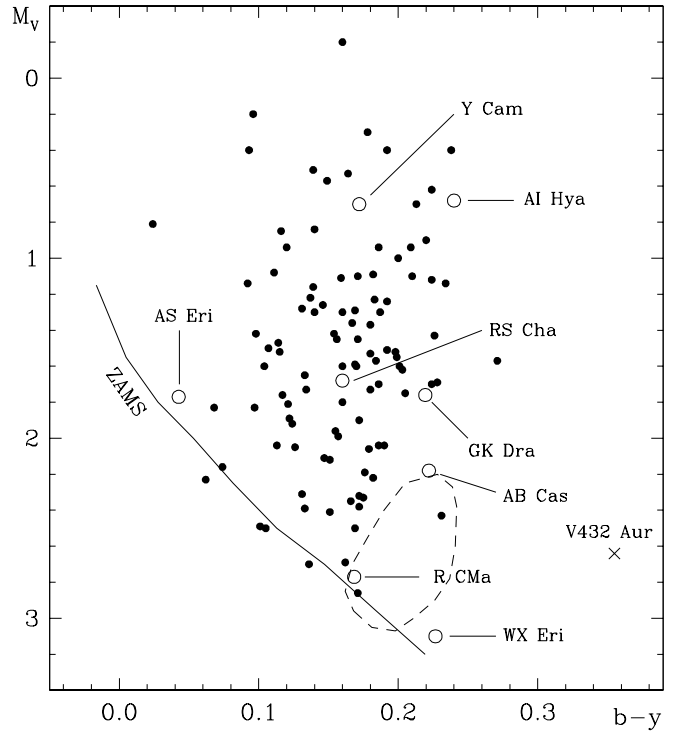,width=8.7cm}}
    \caption{Position on the HR diagram of the $\delta$~Sct component of
     V432~Aur compared with those given by Breger (1979). Position of
     $\delta$~Sct components of eclipsing binaries are plotted with open
     circles and identified (from Broglia \& Marin 1974, Clausen \&
     Nordstr\"om 1980, Griffin \& Boffin 2003, J\o rgensen \& Gr\o nbech
     1978, Popper 1973, Rodr\'{\i}guez et al. 1998, Sarma \& Abhyankar 1979,
     Sarma et al. 1996, Zwitter et al. 2003). The dashed ellipse
     contours the locus of validated $\gamma$~Dor variables according to
     Breger and Beichbuchner (1996) and Kaye et al. (1999).}
    \end{figure}

\section{Comparison with stellar theoretical models}

The theoretical models describe the location of stars on the {\em
theoretical plane} $L,T_{\rm eff}$, while observations place the stars on
the {\em observational plane}, e.g. {\em V, (B-V)}. The correspondence
between them is all but a straightforward one, being strongly affected by
bolometric correction, color-temperature transformation, distance, reddening, 
proper knowledge of the transmission profile of the photometric bands, etc.
The orbital solution in Table~2, providing directly $L$ and $T_{\rm eff}$,
overcomes all such difficulties and allow an assumption-free comparison
between observations and models.

Adopting $T_{\rm eff}$=5770~K and $R$=6.96\,10$^6$~km for the Sun (Allen
1976), the position of the two components of V432~Aur on the $L,T_{\rm eff}$
plane is presented in Figure~3, where a comparison is provided with
evolutionary tracks and isochrones appropriate for the masses (1.22 and 1.08
M$_\odot$) and metallicity ([Z/Z$_\odot$]=$-$0.6) of the system components.
They have been obtained via interpolation over the grid computed by the
Padova theoretical group (Fagotto, et al. 1994, Bertelli et al. 1994,
Girardi et al. 2000, and references therein). Given the high sensibility to
mass and metallicity of the evolutionary tracks, as depicted in the lower
panels of Figure~3, the excellent agreement between observations and theory
is an independent check of the accuracy of the results obtained with the
orbital and atmospheric analysis. The placing of both components exactly
halfway between the isochrones for 3.5 and 4.0 Gyr suggests and age of 3.75
Gyr for V432~Aur.

The secondary star is well evolved away from the main sequence and
approaching the base of the giant branch, while the primary has only
marginally moved away from the ZAMS and is still living its main sequence
phase.

\section{Synchronous rotation and macro-turbulence}

Given the long life lived by both components on the main sequence, there is
little room for doubts that they reached at that time perfect
synchronization between orbital and rotation periods. We can obviously
assume that the rotation axes are aligned with the orbital axis (reached
earlier in the binary evolution according to the time scale discussed by Hut
1981; for deviations from the rule see for ex. Glebocki and Stawikowski
1997). Consequently, from the orbital period and stellar radii in Table~2 it
can be derived that the equatorial rotational velocities for synchronous
rotation at the present time are 20 and 40~km~sec$^{-1}$ for the primary and
the secondary, respectively. The rotational velocities derived by the
atmospheric analysis reported in Table~3 are 20 and 50~km~sec$^{-1}$ for the
primary and the secondary, respectively. While the match is perfect for the
primary, no doubt the line profiles for the secondary star show a broadness
that if interpreted in terms of rotation would indicate the secondary to
rotate {\em faster} than synchronous rotation, even if a precise
quantification at 1~km~sec$^{-1}$ level requires higher resolution spectra
obtained during the totality phase at the primary eclipse. This finding is
unexpected, because one may argue that an {\em expanding} star tends to
rotate slower for obvious arguments based on the conservation of the angular
momentum, and tidal torque forces can bring the star in synchronous rotation
with the orbital motion only if their time scale is shorter than the
expansion time-scale. The time scale for an expansion of 10\% of the radius
of the secondary star of V432~Aur is (from the evolutionary tracks):
\begin{equation}
\Gamma_{(R\uparrow 10\%)} = 6~10^7~~~{\rm yr}
\end{equation}
Following Zahn (1977) the synchronization time for stars with convective envelopes
can be expressed as function of mass-ratio $q$ and period $P$ as:
\begin{center}
\begin{equation}
t_{sync} \approx 10^4 ~\biggl(\frac{1+q}{2q} \biggr)^2 P^4 ~~~{\rm yr}
\end{equation}
\end{center}
or $t_{sync} ~\sim~ 8\, 10^5$ yr in our case, thus 2 orders of magnitude faster
than radius expansion. Therefore, the evolved component of V432~Aur should 
currently be locked into synchronicity. 

The extra broadness of spectral lines of the secondary star is probably not
due to rotation but to other reasons, among which a major contributor could
be a {\em macro-}turbulent velocity field in its atmosphere.  The net effect
of macro-turbulence affects mostly the wings of the lines, making them
broader (cf. Grey 1992). Macro-turbulent motions on a scale larger than
observed with Sun's photospheric granulation have to be expected in the
secondary star of V432~Aur considering that large spots affect its surface
(cf. sect. 3.1) and that it is probably pulsating (cf. sect 7 below).
Macro-turbulent velocities determined by Donati et al. (1995) for the active
surface RS CVn star $\lambda$~Andromedae are $\zeta_{\rm p}$=5.5
km~sec$^{-1}$ for the photosphere and $\zeta_{\rm p}$=10 km~sec$^{-1}$ for
the active surface regions.

Higher resolution observations (with a resolving power not less than
60\,000) during the primary minimum are obviously desirable to better
investigate the line profiles of the secondary star. Higher resolution
observations are required to reduce the $\pm$5~km~sec$^{-1}$ error bar on
the $V_{\rm rot} \sin i$ in Table~4, and to disentangle the pure rotational
broadening of the spectral lines from that of other possible broadening
sources, in particoular the macro-turbulence.

\section{The intrinsic variability of the cooler component}

Da02b already noted how the lightcurve of V432~Aur is affected by an
additional variability superimposed to the eclipsing one. In fact, the light
curves in Figure~1 show a {\em noise} much larger than expected for a 5
millimag-mag precision of the individual points. The flat bottom of the
primary eclipse shows distinct branches corresponding to different observing
epochs, which almost vanish at secondary eclipse. It is therefore clear that
the additional variability is associated with the secondary, larger and
cooler component.

Our photometric observations do not cover sufficiently long, uninterrupted
runs to allow fruitful period search at specific epochs, and analysis of the
whole data set does not provide firm values. The residuals of radial
velocities of the secondary from their orbital model fitting in Figure~1 are
larger than the mean of individual errors (cf. last column of Table~2),
which suggests the presence of a possible pulsation activity. The residuals
have been analyzed with a Deeming-Fourier code (Deeming 1975)
and the resulting power spectrum is presented in Figure~5. Several peaks are
present, with Figure~6 plotting in phase the residuals of radial velocities
according to the strongest peak at 0.1678 day. A pulsation amplitude of the
order of 1.5~km~sec$^{-1}$ seems present. A similar pulsation amplitude is
shown by the multi-epoch radial velocities of the $\delta$~Sct variable
within the GK~Dra eclipsing system (Dallaporta et al. 2002a, Griffin and
Boffin 2003). To estimate the reliability of the 0.1678 day period, we
have randomly removed 15\% of the radial velocity residuals and performed
again the Deeming-Fourier analysis. We have repeated the procedure 50 times,
and 0.1678 day has remained the strongest peak in the power spectrum in 4/5
of the trials. Even if we cannot claim for sure that this is a true
periodicity for V432~Aur, nevertheless it seems the best one with the
available data.

The amplitude of the photometric variability is $\Delta V\approx$ 0.075~mag
and the spectral classification of the intrinsic variable component is
F9\,IV.  According to the review by Feast (1996), the $\delta$~Sct variables
have periods shorter than 0.3 days, their spectral types go from A to F, and
the amplitudes span from a few millimag-mag up to 0.8 mag with a mean around
0.2 mag. The $\delta$~Sct variables may be both radial or non-radial
pulsators, or even display a combination of the two modes, which
superposition causes a beating that makes the lightcurve to appear
frequently chaotic. The shape and amplitude of the light and radial
velocity curves change continuously, as nicely documented by Breger (1977).
Therefore, phase over-plotting of data from a range of different epochs
frequently results in a wash-out of the pulsation shape of the curve, which
appears disappointingly {\em too noisy}. Figure~7 compares the position, on
a color-absolute magnitude diagram, of the secondary star in V432~Aur and of
known $\delta$~Sct stars, highlighting in particular the position of those
contained in eclipsing binary systems. The locus for the $\gamma$~Dor
pulsating variables, frequently discussed in connection to the $\delta$~Sct
ones, is also marked. The secondary star in V432~Aur falls away from both
types of variables, so any association with them is at the moment only
tentative. It is worth to note that Xu et al. (2002) have speculated that a
low metallicity should push the locus of $\delta$~Sct variables on such
diagram toward the position occupied by V432~Aur, which is characterized by
[Fe/H]=$-$0.6 as above derived.

As already stated, our photometric observations are scattered through many
nights, and no single observing run is densely populated and is long enough
to allow to investigate pulsation light curves at any individual epoch. A
long series of consecutive observations does not exist either for the radial
velocities. Firmly establishing if one or more periods are contemporaneously
excited and determining their precise values (and the shape of corresponding
light or radial velocity curves) is mandatory to ascertain the nature of the
intrinsic variability in V432~Aur. To this aim we plan for the coming new
observing season to obtain night-long consecutive series of photometric
(optical and possibly infrared too) and radial velocity observations of
V432~Aur.

\begin{acknowledgements}
We would like to thank R.Barbon for assistance during the whole
project, G.Bono and P.Lampens for useful discussion on $\delta$~Sct stars,
M.Fiorucci for assitance with coding the $\chi^2$ algorithm,
U.Jauregi and F.Castelli for support in handling with Kurucz's spectra, and
C.Boeche who secure a few of the spectra here discussed. R.Sordo was
supported by COFIN-2001 grant.
\end{acknowledgements}

\end{document}